\documentclass{article}
\pdfoutput=1
\PassOptionsToPackage{numbers, compress}{natbib}

\usepackage[final]{neurips_2021_ml4ad}

\usepackage[utf8]{inputenc} 
\usepackage[T1]{fontenc}    
\usepackage{hyperref}       
\usepackage{url}            
\usepackage{booktabs}       
\usepackage{amsfonts}       
\usepackage{nicefrac}       
\usepackage{microtype}      
\usepackage{xcolor}         

\usepackage{lipsum}
\usepackage{wrapfig}
\usepackage{algorithm}
\usepackage{algorithmic}

\usepackage{makecell}

\usepackage{mathtools} 
\usepackage{booktabs} 
\usepackage{tikz} 
\usepackage{dsfont}
\usepackage{amsmath,amssymb}

\usepackage{color}

\title{Incorporating Voice Instructions in Model-Based Reinforcement Learning for Self-Driving Cars}

\author{%
  Mingze Wang \\
  InfoSense, Department of Computer Science\\
  Georgetown University, United States\\
  \texttt{mw1222@georgetown.edu} \\
   \And
   Ziyang Zhang \\
   InfoSense, Department of Computer Science\\
   Georgetown University, United States \\
   \texttt{zz249@georgetown.edu} \\
   \AND
   Grace Hui Yang \\
   InfoSense, Department of Computer Science \\
   Georgetown University, United States \\
   \texttt{huiyang@cs.georgetown.edu} \\
}

\begin{document}

\maketitle

\begin{abstract}

This paper presents a novel approach that supports natural language voice instructions to guide deep reinforcement learning (DRL) algorithms when training self-driving cars. DRL methods are popular approaches for autonomous vehicle (AV) agents. However, most existing methods are sample- and time-inefficient and lack a natural communication channel with the human expert. In this paper, how new human drivers learn from human coaches motivates us to study new ways of human-in-the-loop learning and a more natural and approachable training interface for the agents. We propose incorporating natural language voice instructions (NLI) in model-based deep reinforcement learning to train self-driving cars. We evaluate the proposed method together with a few state-of-the-art DRL methods in the CARLA simulator. The results show that NLI can help ease the training process and significantly boost the agents' learning speed.
\end{abstract}

\section{Introduction}

Research in self-driving cars, also known as autonomous vehicles (AV),  has made rapid progress due to significant advancements in deep learning. 
Deep reinforcement learning (DRL) methods are popular approaches for AV. A DRL agent learns by interacting with the driving environment and gradually forms a good policy that satisfies constraints in the  environment. An effective policy can take up many trial-and-error learning experiences. In each learning episode, the agent improves its policy slightly towards an optimization objective, which in most cases is to  maximize the long-term expected cumulative rewards~\citep{rl}.
While these algorithms show promising results, most of them are sample inefficient and require substantial training episodes.
Model-based reinforcement learning has been proposed to improve sample efficiency by learning an environmental model and using the model to better plan the actions. 

However, learning the policy and that of the environmental model is still quite time and sample-consuming.  
When watching an agent's learning curve develop, it is not uncommon for an engineer to have certain words or thoughts come to her mind and almost verbalize to the agent: ``Hey, that move is stupid'', ``Good job!'' or ``Just go there.'' When she sees surprisingly good moves, she would like to let the agent know, and vice versa. However, such a ``direct'', natural, and quick communication channel does not exist.  

On the other hand, we notice that a human novice driver could successfully learn driving with much less training. Human drivers have a human coach giving instructions and evaluative feedback. During a lesson, the student driver does the driving, the coach verbally delivers commands, praises, scolds, and sometimes explains traffic rules and driving conventions paired with stories and reasoning. Their way of communication is natural language (NL) utterances. The student may not speak much while focusing on driving and learning; the coach does most of the talking. Compared to an artificial learner, only a small amount of instructions from the coach is needed by the human student and she learns quite efficiently. On average, a new human driver takes three to five behind-the-wheel classes, each of which lasts around two hours, to pass the road test for a class-D license.\footnote{\url{https://dmv.dc.gov/page/mandatory-driver-education}.}

This motivates us to propose a new way to communicate and train autonomous car agents. In this paper, we propose to give natural language voice instructions to train AV agents directly. Compared to keyboard strokes or other control commands, natural language is a very convenient, natural, and fast communication method. Recent progress in Automated Speech Recognition (ASR), Natural Language Processing (NLP), and personal assistants should have been proliferated to allow people to interact with the learning agents with their voice naturally. Our goal is to provide the agents with the most natural and direct interference and improve and accelerate their learning. 

The voice training interface that we present in this paper can potentially facilitate two scenarios. First, it can help machine learning practitioners interact with their learning algorithms more naturally and directly impact the agent's learning process. Second, the voice inputs allow the non-technical end-user to continue the training after purchasing the agent home. It is like online learning after they  have owned the product for personal use. 

In this paper, we propose incorporating voice instructions in model-based deep reinforcement learning to train self-driving cars. First, we collect voice instructions offline and build a training dataset and a taxonomy of NL instructions. The data analysis shows that most NL instructions are indeed about actions and rewards for the agent. Second, we build a human-in-the-loop model-based DRL framework that can incorporate live voice instructions. Two places in a model-based RL algorithm can incorporate natural language instructions. The first is policy learning, where the natural language instructions can improve or overwrite the agent's actions. The second is the environmental reward model. Ideally, a reward function for the self-driving car would need to handle multiple objectives, such as safety, conforming to traffic rules, speed under control, and passengers feeling comfortable. 
Oftentimes, the reward function is pre-set and hand-crafted by experts before the learning starts, which is not flexible and may not adapt well to the dynamically changing situations while driving. Allowing human interference during training can diversify the reward function from its pre-set formulation and better suit a real-world situation. 

\begin{figure}[t]
\begin{minipage}[b]{\textwidth}
    \begin{minipage}[t]{0.44\textwidth}
        \centering
        \includegraphics[width=0.9\linewidth]{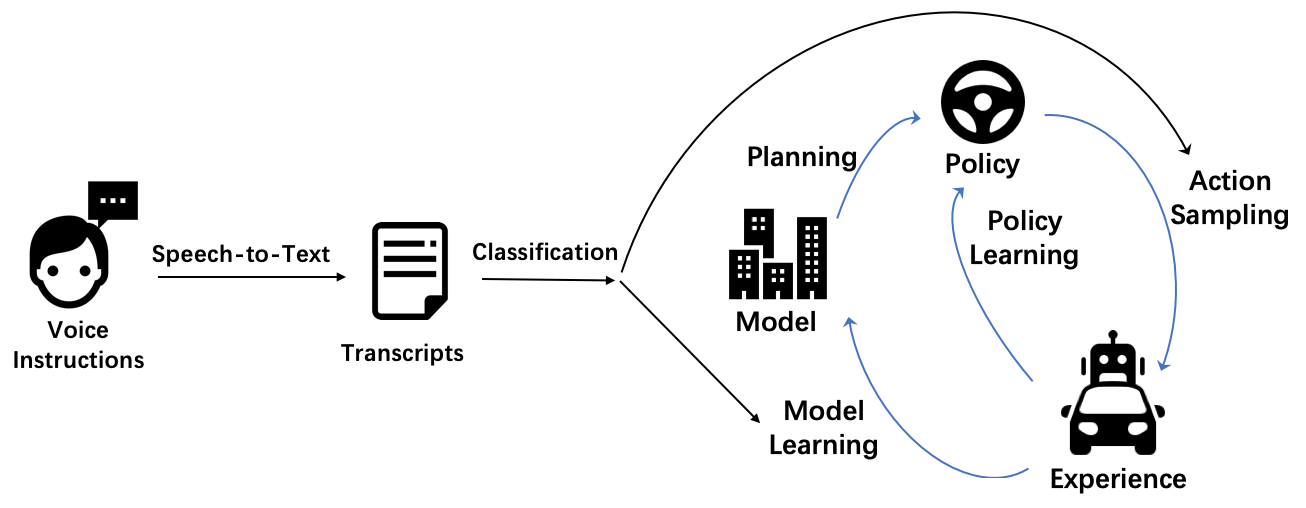}
        \caption{Model-Based RL with Voice Instructions.}
        \label{structure}
    \end{minipage}
    \hspace{10pt}
    \begin{minipage}[t]{0.19\textwidth}
        \centering
        \includegraphics[width=0.8\linewidth]{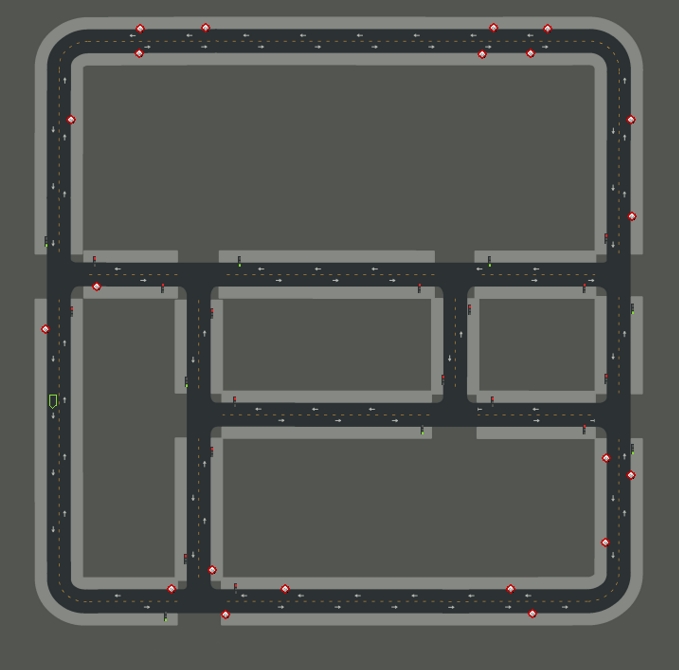}
        \caption{Example town in CARLA}
        \label{fig:map}
    \end{minipage}\hfill
    \begin{minipage}[t]{0.32\textwidth}
        \centering
        \includegraphics[width=0.95\linewidth]{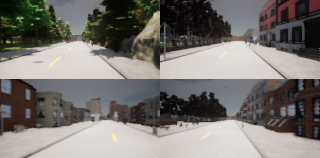}
        \caption{Front camera inputs}
        \label{fig:front_camera}
    \end{minipage}\hfill
\end{minipage}
\end{figure}

We evaluate our method in the CARLA simulator~\citep{carla}. The simulator's image resolution is lower than reality and has fewer traffic signs (only supports traffic lights, stop signs, and speed limits). Moreover, 
the simulated town is small and other vehicles drive idealized, without any possible  aggressive drivers. However, the open-source simulator supports flexible sensor suites, environmental conditions, full control of all static and dynamic actors, map generation, and many other functionalities. One of the most popular simulator~\citep{chen2019learning,prakash2020exploring}s, CARLA is sufficient to test our idea.

Our method is general and can be used in combination with various learning methods for self-driving cars. Particularly, we investigate Dyna~\citep{rl} with PPO~\citep{ppo}. Experiments show that using voice instructions to interfere with self-driving cars' learning can significantly improve the learning efficiency and make self-driving agents converge much faster to a good policy.

\section{Related Work}

\textbf{Research in Autonomous Vehicles.}
Research on autonomous driving agents can be categorized as modular models and end-to-end models~\citep{dlforad}.
Traditionally, self-driving cars use a pipeline of perception-planning-action modular models to compute their driving decisions~\citep{driveinday}. More recently,  end-to-end learning methods, which directly map from raw sensor inputs to control commands~\citep{dlforad}, have become popular for autonomous driving. Many are deep reinforcement learning (DRL) methods that directly optimize the whole system's effectiveness from raw inputs and environmental rewards. For instance, \citep{race} asynchronous actor-critic agents (A3C)~\citep{a3c} were used to map image inputs from a forward-facing camera to control a car agent in a game simulator. The algorithm uses millions of interactions with the environment to converge to a good racing policy. \citep{driveinday} employed deep deterministic policy gradient~\citep{ddpg} for their self-driving car agents. Their agent learns to follow a 250m rural lane within only half an hour of training. 
\citep{carlappo} applied proximal policy optimization (PPO) \citep{ppo} to a lane following task in the simulator CARLA, with an average of eight hours of training. All three pieces of work can solve the simple lane following problem, where the car drives on a route with no intersections, no traffic lights, and no other vehicles. However, even with such a simple task setting, the training process is still largely inefficient, especially at the beginning of the training.

\textbf{Simulated training.} It can be an extremely high cost to train self-driving cars on real roads. Therefore, many algorithms use simulators to train the agents. CARLA~\citep{carla} is a popular simulator used in recent research~\citep{chen2019learning,prakash2020exploring} for autonomous driving research. It can support the development, training, and validation of autonomous driving systems. The simulator provides a flexible specification of sensor suites, environmental conditions, full control of all static and dynamic actors, map generation, and many other functionalities.
As a simulator, CARLA's limitations are mostly its simplification to real-world situations. For example, it has limited types of traffic signs, low texture resolution, and small towns. But for the essential tasks of a self-driving car, such as following the traffic lanes and watching out the pedestrians, CARLA is sufficient. This work uses the ``Town02'' environment  with random weather settings and a limited fresh rate of 10 per second.

\textbf{Human-in-the-loop learning.} Our work belongs to human-in-the-loop learning (HILL), a recent machine learning paradigm that has generated significant practical and theoretical interests~\citep{9131765}.
We group HILL research into the following categories by the {\it usages of the human interventions} and by the {\it types of human feedback}. 

Based on where an RL algorithm uses human intervention, we can classify the works into direct policy learning, reward learning, and policy and reward learning. The first and most common usage of human intervention is providing evaluative feedback directly to the action generated by a policy~\citep{arumugam2019deep,macglashan2017interactive}. These methods are simple and effective. They help the policy learning quickly converge. The second type is to incorporate human feedback in reward learning, aka inverse reinforcement learning. The idea is to use human trainers to control the robots or agents to generate a demonstration~\citep{argall2009survey,article} and the inverse reinforcement learning can benefit from the generated demonstrations to learn the reward functions~\citep{10.1145/1015330.1015430}. 

Various types of human interventions have been explored in HILL.
For example, the simplest one is the hardware delivered feedback, mainly including mouse clicks, keyboard keys, and other sensors~\citep{knox2009interactively,macglashan2017interactive}. Although this method is precise, it needs reaction time for human trainers to express the feedback by using a keyboard or mouse. Instead, we can use mores natural ways to provide feedback such as facial feedback~\citep{arakawa2018dqntamer,broekens2007emotion}, natural language feedback~\citep{Kuhlmann2004Guiding,maclin1996creating} and gesture feedback~\citep{cruz2016multi,yanik2013gesture}. VNLA~\citep{nguyen2019visionbased} use natural language in a navigation task. If the agent is stuck somewhere or gets lost, the human trainers give the agent detailed instructions on reaching the next sub-goal. The instructions contain the direction and the features of the sub-goal, which is easy for the agent to detect and reach. However, the natural language feedback in VNLA is not in real-time. For AV tasks, real-time instructions are more effective than non-real-time ones. Our approach uses real-time instructions to coach the agents on driving and achieve higher AV tasks.

The most similar piece of work to ours is perhaps the Deep COACH (COnvergent Actor-Critic by Humans) method~\citep{arumugam2019deep}. Deep COACH works with a human trainer as part of the deep actor-critic reinforcement learning algorithm. It considers human feedback as a more accurate evaluation of the agent's choice than a reward function. The human trainer acts as the critic, and the objective policy acts as the actor. Thus the relationship between human trainers and the objective policy is suitable for the actor-critic algorithm. The types of human interventions used in Deep COACH are simple positive or negative values. However, an actual human trainer can indeed provide more informative information to the agent beyond simple positive or negative feedback. In this paper, we propose to use natural language as the communication channel between the human trainer and the robotic car agent, aiming to provide the agents with richer and more precise instructions that make full use of human trainers.

\begin{table*}[t]
\centering 
\caption{Taxonomy of Voice Instructions.}
\label{ins} \small
\resizebox{!}{0.4\textwidth}{
\begin{tabular}{llll} 
\toprule
Type     & Sub-type  & Example      &  Implementation \\
\midrule
Action & Go Straight & Just keep your car straight.  & $steer = 0$ \\
       &  & Hold steering wheel steady.   & \\
       &  & That's too much oscillation. & \\    
       & Turn Left & To the left.& $steer = -0.2$ \\
       &  & Take left, take left, left. & \\
       & Turn Right & Take a right.& $steer = 0.2$ \\
       &  & Move, right. & \\
       & Speed up & You can probably speed up. & $throttle += 0.15$ \\
       &  & Go go, a bit faster. & \\
       & Slow down & oming up slow, slow down. & $throttle -= 0.15$ \\
       &  & You need to slow down. & \\
       & Stop & Stop sign.  & $brake = 1$  \\ 
       &  & Stop! &  \\ 
       &  & There's a stop sign so we have to stop. &  \\ 
\midrule 
Reward & Great & Cool! & $r_h = 30$ \\
       &  & Yeah, you are doing well. & \\
       & Good & All right. It's ok. & $r_h = 10$ \\
       & Bad & Oh, no. & $r_h = -10$ \\   
       &  & That's that's not good. & \\
       & Terrible & What are you doing? & $r_h = -30$ \\  
       &  & You can't do that. & \\
       & Mistake & You blew another red light.  & $r_h = -30$ \\
       &  & You're on the sidewalk.  &  \\
       & Accident & You ran into a wall. & $r_h = -30$ \\
       &  & You blew a stop sign. & \\
\midrule 
Reasoning & Causal & There's a stop sign so we have to stop. &  \\  
 & &You're approaching a yield. So you may want to look for cars. & \\
 && No need to stop. You. Didn't need to stop there. There's no sign.  &\\
\midrule
State & Describe  & There's a stop sign. &  \\  
 & &You're at a red light. & \\
 && Watch out the grass!  &\\
\bottomrule
\end{tabular}
}
\end{table*}

\section{Constructing a Voice Instruction Taxonomy}

Our research aims to give natural language voice instructions directly to self-driving car agents for more effective and efficient training. The challenge of this project is to train the agent to understand natural language coaching instructions. To serve this goal, we collected a training dataset and built an NL instruction taxonomy from it. The dataset has been annotated and will be made available for research use upon publication. 

\textbf{Collecting Natural Language Instructions.} To build the dataset, we gather audio recordings from machine learning practitioners for their voice instructions when they are watching the agents' driving. Our goal is to create a representative, natural, free-form voice instruction dataset. 

The participants were instructed to imagine self-driving car coaches teaching a new driver (the self-driving car) in a driving school. Their instructions to the agent should be natural conversations instead of from a controlled vocabulary. For instance, the instructions can be direct commands, praises, curses, or rules that they wish the agent to learn. The participants were encouraged to be creative and talk freely, as long as the instructions fit the role of a driving coach. Six participants who are familiar with machine learning and driving participated in the data collection. The participants' names or user ids are fully anonymized and the data are password-protected and stored in a secure place. The participants were also instructed to only talk in the context of teaching an agent to drive and not to talk about personal, medical, or other sensitive information. 

Four types of sample driving videos are provided based on their training levels: 
(1) ``Non-trained'' are videos from a random self-driving car agent who receives no training at all. It can act quite weird and full of crashes. (2) ``Mid-trained'' are videos from self-driving car agents whose learning curves are not yet converged and only half-trained. They can follow the lane twisty. (3) ``Trained'' are the videos from the self-driving car agents who have been full-trained by a model-free RL method and can follow the lane correctly. (4) ``Autopilot'' are videos produced by the autopilot control in the simulator. It can be considered as the gold standard that an artificial agent can achieve. When watching the videos, the participants were instructed to record their voice instructions using an audio recording device. They were also asked to try their best to synchronize the voice to the video. In the end, we collected 20 audio clips. 

\textbf{Speech-to-Text Transcription.} We used Google's speech-to-text (STT)~\citep{google_STT} to convert these audio clips into text transcripts. The text transcripts are then segmented into utterances and aligned with the videos. The utterances range from short phrase-like commands to long sentences. The content includes direct commands such as ``go straight'', positive recognition, expressions of surprises and disappointment, and reasoning. These utterances may not be 100\% grammatically correct. However, they show us a representative group of expressions that a human coach would use to teach a car agent.  

Note that the speech-to-text tool can also perform real-time speech recognition. Each time the microphone receives an instruction, it sends it to the server of STT and returns the corresponding text scripts quickly. We present our use of it for real-time transcription in later sections. In addition, we are also aware that audios contain more information than their text transcripts. For instance, intonations, volume level, exclamations, and signs would be important signals to express evaluative feedback to the agents or infer the human's real intention. However, we leave these to future work and concentrate on textual analysis of the dataset. 

\textbf{Building Taxonomy of Instructions.} According to the real-world driving experience, we analyze and semi-automatically organize the voice instruction dataset into a taxonomy. Table~\ref{ins} shows our two-level taxonomy, examples and how to implement them into the learning algorithm.

The two most common types of instructions in the utterances that we collected are actions and rewards. Actions mean the instructions tell the agent how to drive in the next step. In this type of instructions, we have steer commands {\it turn right}, {\it turn left} and {\it go straight} and speed commands {\it speed up}, {\it speed down} and {\it stop}. The reward-related instructions give evaluative feedback to the agent. Some of them indicate severity of an issue, so we use them as different reward levels: {\it great}, {\it good}, {\it bad} and {\it terrible}. Others directly mention the event type, for instance, {\it mistake} or {\it accident}.

\section{Model-Based RL with Voice Instructions}

This paper proposes incorporating natural language voice instructions in model-based reinforcement learning to train self-driving car agents. Our goal is to support more natural and direct interference from human engineers or end-users to the agents and improve and accelerate the agents' learning. 

\textbf{Model-Based Deep RL.} Similar to most reinforcement learning methods, model-based reinforcement learning is built on top of the Markov Decision Process (MDP). It consists of a tuple $\langle \mathbb{S}, \mathbb{A}, \mathbb{R}, \mathbb{T} \rangle$ to represent state, action, reward, and state transition function. 
\begin{itemize} 
\item State: $s \in \mathbb{S}$, consisting two parts $s=(i, v)$,  $i \in \mathbb{I}$, where $\mathbb{I}$ is the set of all possible images captured from cameras, and $v \in \mathbb{R^+}$, the current car running speed. The camera image set $\mathbb{I}$ can include front, side, and rearview. But in this work, we only use the front view, which is  adequate for most tasks in CARLA. 
\item Action:  $a \in \mathbb{A}$, consisting of steer $a_{steer} \in [-1,1]$, throttle $a_{throttle} \in [0,1]$ and brake $a_{brake} \in [0,1]$, represented by a vector of length three $a = [a_{steer}, a_{throttle}, a_{brake}]^T$.
\item Reward function $\mathbb{R}$: part of the environmental model in a model-based RL algorithm and is learned via model learning. 
\item Transition function $\mathbb{T}$: also part of the environmental model and is learned via model learning. However, it is not in the interest of this paper and we will skip it.  
\end{itemize} 

Model-based reinforcement learning has two concurrent learning components. One is similar to model-free reinforcement learning, known as ``policy learning'' or ``direct RL.'' Policy learning's goal is to find the best policy $\pi^*$ directly via environmental rewards, where assuming the transition model and the reward function are deterministic. Policy learning aims to maximize the objective function $J(\pi)$, which is the expected value of the cumulative future reward $E_{\pi} \left[ \sum_{k=t}^{T} \gamma^k R(s_k, \pi(s_k)) \right]$. Model learning is the other learning component. In our work, we focus more on reward function learning because the CARLA simulator decides the environmental state transition. 

In this work, we propose incorporating voice instructions into both the model-based RL's policy learning and model learning. Next, we first present how we obtain and process the voice inputs in real-time, followed by details of incorporating the processed instructions in policy learning and model learning, respectively. The overall algorithm is presented in Algorithm 1.

\begin{figure}[t]
    \begin{algorithm}[H]
        \caption{Teaching AV Agents with Voice Instructions.}\label{alg:highlevel}
    \algsetup{linenosize=\small}
    \small
    \begin{algorithmic}[1]
        \STATE {\bfseries Input:} training environment $E$; voice action $a_h$, voice reward $r_h$; number of trajectories each epoch $N$; replay buffer $B$; reward coefficients $\beta$, policy $\pi_{theta}$ with initial parameters $\theta$ 
        \REPEAT
        \FOR{$t = 0$ {\bfseries to} $N$}
        \STATE initialize $done$, $traj$, $E$
        \WHILE{not $done$}
        \STATE Get current state $s$ from $E$.
        \STATE Sample action $a$ from $\hat{\pi}_n$
        \IF{voice action $a_h \neq \emptyset$}
        \STATE $a = a_h$
        \ENDIF 
        \STATE Get reward $r$ and next state $s'$ after executing $a$
        \IF{voice reward $r_h \neq \emptyset$}
        \STATE Update $\beta$ for an earlier reward $r^{prev}$ with $r_h$
        \FOR{$i = prev$ {\bfseries to} $t-1$}
        \STATE Update $(s, a, r)$ at timestep $i$ in $traj$ based on $\beta$ 
        \ENDFOR
        \ENDIF
        \STATE Append $(s, a, r)$ to $traj$
        \ENDWHILE
        \STATE $B$ stores $traj$
        \ENDFOR
        \STATE Update $\hat{\pi}_n$ with $(s, a, r)$ in buffer $B$
        \UNTIL{Algorithm converges.}
    \end{algorithmic}
    \end{algorithm} 
\end{figure}

\textbf{Mapping Voice Instructions to Actions and Rewards.} Our project aims to use human instructions in real-time to teach a learning car agent. One option is to train an end-to-end neural network classifier to map raw audio clips to our instruction categories (Table \ref{ins}). However, this could be impractical due to the live interaction constraint. We thus propose a two-stage process that first performs speech-to-text transcription over the speech data and then classifies the textual utterances into  action- or reward-related commands that the agent can easily operate. 

First, we use Google's speech-to-text (STT) tool~\citep{google_STT}, which can convert audio into text with real-time speech recognition. Each time the microphone receives voice instructions, it sends it to the server of STT and returns the corresponding text quickly. This service enables us to process a human coach's live instructions like in a behind-the-wheel driving class. The accuracy of STT depends on many factors, such as speech tones, accent, and microphone quality. The accuracy of STT reported by Google Team is 93.1\% (6.9\% Word Error Rate)~\citep{prabhavalkar2017minimum}.

Second, we employ a BERT-based neural network text classifier~\citep{devlin2019bert} to classify each transcribed utterance, i.e., a verbal sentence that the human trainer spoke, to one of the actions or reward sub-types. BERT is a general-purpose and robust pre-trained language model widely used in natural language processing (NLP) tasks. We mainly use BERT-Base's model~\citep{turc2019wellread} to translate the sentences into vectors. The vectors are then fed to a three-layer multi-layer perceptron (MLP) network to make the final classification decision. Our method can reach 94.4\% accuracy in the experiments.

\textbf{NL Instructions for Policy Learning.} The neural network classifier maps a real-time NL instruction $ins$ to action $a_{h}$ in the agent's action space. We consider the new action command from the human coach $a_{h}$ has the highest priority so that it can overwrite the agent's sampled action generated from the agent's policy. In Algorithm~\ref{alg:highlevel}, at each interaction loop, if a piece of instruction $ins$ comes, the agent replaces the sampled action $a$ with $a_h$ for $t_d$ seconds. 

The idea of replacing the agent's actions with $a_h$ intuitively works with RL because the human coach's policy, once executed, would be reinforced if it leads to higher rewards. 

When using the original PPO \cite{ppo} as the direct policy learner, one issue arises that after the sampled actions are replaced with $a_h$, the ratio $\eta(\theta)$ becomes $\frac{\pi_{\theta}(a_h|s_t)}{\pi_{\theta_{old}}(a_h|s_t)}$. Note that NL instruction $a_h$ is not sampled from the agent’s policy. Thus, it may drastically change the agent's policy (if the human trainer's policy deviates far from the agent’s policy) and the policy may not be able to converge. We propose to modify PPO's regularization by changing the clipping function. Our solution is to clip the old probability by a fix number: $\eta(\theta) = \frac{\pi_{\theta}(a_h|s_t)}{max\{\pi_{\theta_{old}}(a_h|s_t), 0.05\}}$. Empirically. we find it helpful to clip the old probability to be larger than 0.05.

\textbf{NL Instructions for Model Learning.} Our reward function $r$ is designed similarly to~\citep{carlappo}'s and based on three factors: speed, deviation, and orientation. For a non-terminal state, the reward function is defined as the multiplication of all three factors. For a terminal stats, it uses a penalty  score $r_T$:
\begin{equation} 
\label{eq4}
  r =
\begin{cases}
    \log \exp \sum \beta_i r_i & \text{Non-terminal states} \\ 
    r_{T}             & \text{Terminal states} \\
\end{cases}, 
\end{equation} 
where $r_i$ is one of $r_s, r_o, r_d$ and $\beta_i$ is the weights to each reward component. The {\it speed} factor $r_{s}$ is a function of the car speed $s: r_{s} (s) = \mathds{1}_{\{0 \leq s < s_{min}\}} \times \frac{s}{s_{min}} + \mathds{1}_{\{s_{min} \leq s < s_{target}\}} + \mathds{1}_{\{s_{target} \leq s < +\infty\}} \times (1 - \frac{s - s_{target}}{s_{max} - s_{target}})$, 
where $s_{min} =\ 10km/h, \ s_{target} =\ 15km/h, \ s_{max} =\ 20km/h$. The {\it deviation} factor $r_{d}$ is a function of the distance $d$ between the center of the car and the center of the road: $r_{d}(d) = max \{1-\frac{d}{d_{max}}, 0\}$, 
where $d_{max}=2.5m$. 
The {\it orientation} factor $r_{o}$ is a function of the angle $a$ between the orientation of the car and the orientation of the road:
$r_{o}(a) = max \{1-\frac{a}{a_{max}}, 0\}$, where $a_{max} = 20^{\circ}$. The self-driving car reaches a terminal state if it drives off track ($r_T = -50$) or collides with any other object ($r_T = -100$) or stops moving for a period of time ($r_T = 0$).

The NL voice reward $r_{h}$ gives evaluative feedback to the agent. After receiving it, we use $r_{h}$ to update a previous state's reward function instead of the current state's. This is because of the latency of proceeding with the human voice input in real-time. According the frame rate $fps$ and pre-set latency $lat$, we find the previous state $s^{t-g}$ based on the time gap $g = t - \frac{lat}{fps}$ between the current state $s^t$ and $s^{t-g}$. We then use supervised regression to learn the reward parameters at $t-g$ that minimize the cost to $r_{h}^t$ at $t$ : \begin{equation}
    \beta*^{t-g} = \arg\min_\beta \sum || r^{t-g},  r_{h}^t||, 
\end{equation}
The newly learned reward function would then be used to update the time steps forward until updated by another human input.

\section{Experimental Results}

\begin{table*}[t]
    \centering \small
    \caption{Effectiveness in tasks without other vehicles and pedestrians. The full lap task tests if an agent can drive stably in a lap. }\label{drl_baseline}
    \resizebox{!}{0.105\textwidth}{
    \begin{tabular}{|l|c|r|r|r|}
    \hline
     & Lane following  & \multicolumn{3}{|c|}{Full Lap} \\
    \hline
     & Success rate & Lap Complete & Cum. Rw & Avg Dev. \\
    \hline
    \makecell[l]{DQN~\citep{dqn}} & 17 & 13.61 (std 6.96) & 81.19 (47.13) & 1.01 (0.24) \\
    \makecell[l]{SAC~\citep{sac}} & 57 & 15.91 (8.05) & 90.72 (84.10) & 0.84 (0.17) \\
    \makecell[l]{DA-RB+~\citep{prakash2020exploring}} & 45 & 1.98 (0.94) & 6.32 (3.28) & 13.65 (5.25) \\
    \makecell[l]{DDQ~\citep{pmlr-v48-gu16}}  & 55 & 26.29 (5.40) & 181.61 (55.78) & 1.1 (1.02) \\
   \hline
    \makecell[l]{DynaPPO~\citep{carlappo,rl}} & 60 & 18.65 (6.89) & 272.87 (131.68) & 0.5 (0.07) \\
    \makecell[l]{DynaPPO+NLI (this paper)} & {\bfseries 100} & {\bfseries 42.95} (10.24) & {\bfseries 635.96} (170.89) & {\bfseries 0.4} (0.03) \\
    \hline
    \makecell[l]{Deep COACH~\citep{arumugam2019deep}} & 21 & 21.22 (5.30) & 82.09 (14.32) & 0.61 (0.03) \\
    \hline
    \end{tabular}
    }
    \centering 
    \small
    \caption{Effectiveness in tasks with vehicles and pedestrians. The avoidance task tests whether the agent can correctly avoid crashing with other vehicles and pedestrians and follow the lane simultaneously.}\label{drl_baseline_avoid}
    \resizebox{!}{0.105\textwidth}{
    \begin{tabular}{|l|@{\hspace{12pt}}c@{\hspace{12pt}}|r|r|r|}
    \hline
     & Avoidance & \multicolumn{3}{|c|}{Full Lap (Avoidance)} \\
    \hline
     & Success rate & Lap Complete & Cum. Rw & Avg Dev. \\
    \hline
    \makecell[l]{DQN~\citep{dqn}} & 7 & 3.61 (std 5.77) & 65.63 (107.20) & 0.373 (0.158) \\ 
    \makecell[l]{SAC~\citep{sac}} & 40 & 12.71 (8.45) & 142.31 (100.10) & 0.8 (0.24) \\ 
    \makecell[l]{DA-RB+~\citep{prakash2020exploring}} & 27 & 0.89 (2.67) & -7.94 (6.16) & 32.66 (9.49) \\
    \makecell[l]{DDQ~\citep{pmlr-v48-gu16}} & 40 & 17.49 (8.79) & 121.22 (57.00) & {1.01} (0.77) \\
    \hline
    \makecell[l]{DynaPPO~\citep{carlappo,rl}} & 57 & 16.85 (8.52) & 266.80 (163.10) & 0.71 (0.20) \\
    \makecell[l]{DynaPPO+NLI (this paper)} & {\bfseries 73} & {\bfseries 18.88} (6.40) & {\bfseries 296.55} (113.65)  & {\bfseries 0.64} (0.14) \\
        \hline
   \makecell[l]{Deep COACH~\citep{arumugam2019deep}} & 13 & 6.41 (4.94) & 30.84 (61.33) & 0.75 (0.18) \\
       \hline
    \end{tabular} 
    }
\end{table*}

We evaluate our method on the CARLA simulator~\citep{carla}, which is a popular open-source simulator for autonomous driving cars. 
The CARLA tasks we experiment on include the {\it Lane following} task, {\it red lights and stop signs} task, {\it full lap} task, and {\it avoidance} task. We also test in both situations, either with an empty street or with other vehicles running and pedestrians walking in the street. The map of Town02 is used, with random weather initiated. We limit the frame rate to 10 per second and use the asynchronous model. 

\subsection{Experimental Setup}
\textbf{Metrics.} For evaluation, the following metrics are measured. 1) The accumulative reward $G(\pi)$ of each episode: $ G(\pi) = \sum_{k=1}^{T} \gamma^k R(s_k, \pi(s_k))$ 2) The success rate out of 100 times for each task. In the lane following task, a whole episode without driving out of the lane is considered a success. In the avoidance task, finishing a whole episode without crashing into other vehicles or pedestrians is considered a success. The final success rate $Success Rate = \frac{\# \text{of success episodes}}{100}$. 3) The completion percentage $P$: $P = \frac{Length_c}{Length} \times 100\%$, where $Length$ represents the total length of the testing lap and $Length_c$ represents the length of the lap the self-driving car goes through before the episode terminates. 4) Lane Deviation: Lane deviation is the distance between the car and the center of the lane. It measures the self-driving car's ability to drive on its lane precisely.

\textbf{Baselines.} To show the effectiveness of our method, we compare  with several best-performing model-free RL methods such as DQN~\citep{dqn}, PPO~\citep{carlappo}, SAC~\citep{sac} and model-based RL baseline methods such as Dyna (with PPO) and DDQ~\citep{pmlr-v48-gu16}. DA-RB+~\citep{prakash2020exploring} is the SOTA for CARLA at the moment and it is an imitation learning method. We report our implementation of it here. 
In addition, we compare with a human-in-the-loop learning method, Deep COACH~\citep{arumugam2019deep}. These methods share the same training interaction experiences and use similar convolutional neural network architectures for image feature extraction.
\textbf{Implementation Details.} We implement the above algorithms in Tensorflow and CARLA 0.9.10. We use a five-layer residual CNN to process the image input, and the input size is $160\times80$ RGB. After five convolution layers, the network receives the speed $v$ and combines it with CNN's output. After another two dense layers, it returns the final result.

The action sample rate is $10HZ$, which means that each interaction step takes 0.1 seconds.

\subsection{Effectiveness} 

We compare the proposed algorithms with state-of-the-art self-driving car systems in  different fields such as model-based learning, model-free learning and HILL. The results are shown in Table~\ref{drl_baseline} and Table~\ref{drl_baseline_avoid}. Table~\ref{drl_baseline} shows the performance of AV agents without the interference of other vehicles and pedestrians. Our method outperforms the baseline models and reaches a 100\% successful rate in lane following task while the best result of other models is 57\%. In the full lap task, our agent drives 63\% further than others and consequently reaches a higher cumulative reward. Also we have the lowest average deviation of 0.4, which means our agent drives are more stable than baselines. This is because the baseline agents are continuously changing steer angle. Eventually they follow the lane but they cannot find that the best solution is holding the steer and driving straight. With NLI guidance, we can let the agent drive straight which leads to a higher cumulative reward. After training, the agent can finally learn not to change steer frequently and finally reaches lower deviation.

Table~\ref{drl_baseline_avoid} shows the performance when we add pedestrians and vehicles to the simulator. All methods have lower performance compared with Table~\ref{drl_baseline}. This is reasonable because the pedestrians can suddenly cross the street and the other vehicles may overtake you and cut your lane. These add many unstable factors to the environment and agent planning policy. For the human, we are taught to stop better when meeting these situations. However, the agent cannot learn it easily. Therefore, we find that the baseline methods may take a sharp turn to avoid the pedestrians but lose control soon. But we can ask the agent to stop in front of pedestrians and vehicles when training. Thus, we have better performance than the baselines.

\begin{figure}[t]
    \begin{minipage}[t]{0.33\textwidth}
        \centering
        \includegraphics[width=0.98\linewidth]{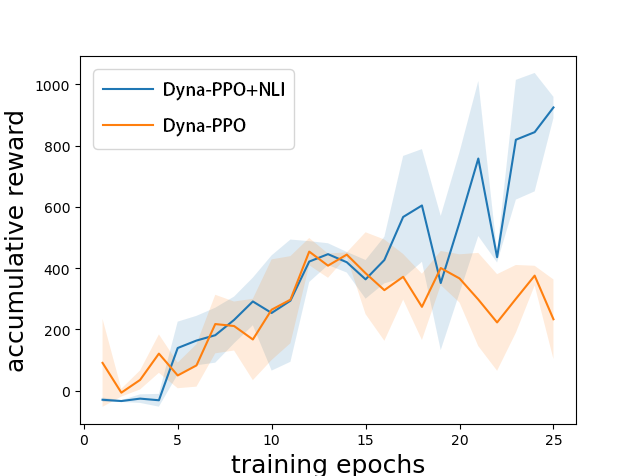}
        \label{fig:lane_accr}
    \end{minipage}\hfill
    \begin{minipage}[t]{0.33\textwidth}
        \centering
        \includegraphics[width=0.98\linewidth]{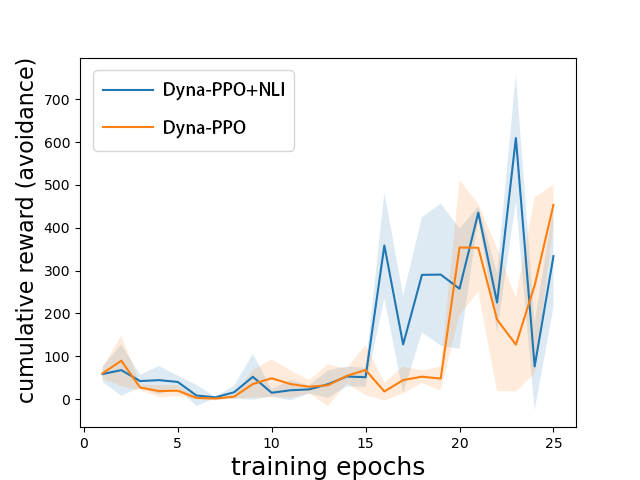}
        \label{fig:avoi_accr}
    \end{minipage}\hfill
    \begin{minipage}[t]{0.33\textwidth}
        \centering
        \includegraphics[width=0.98\linewidth]{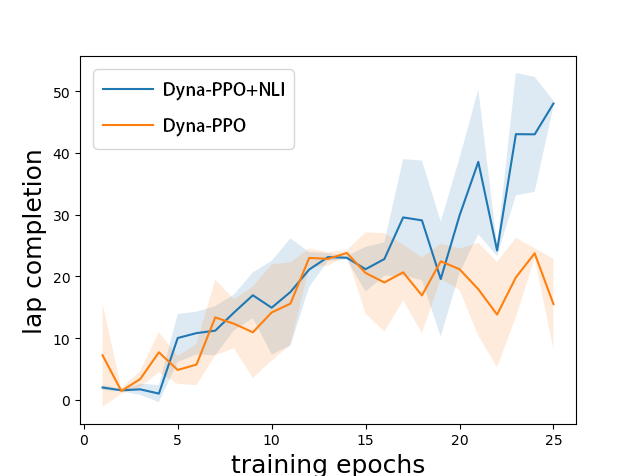}
        \label{fig:lane_comp}
    \end{minipage}\hfill
    \begin{minipage}[t]{0.33\textwidth}
        \centering
        \includegraphics[width=0.98\linewidth]{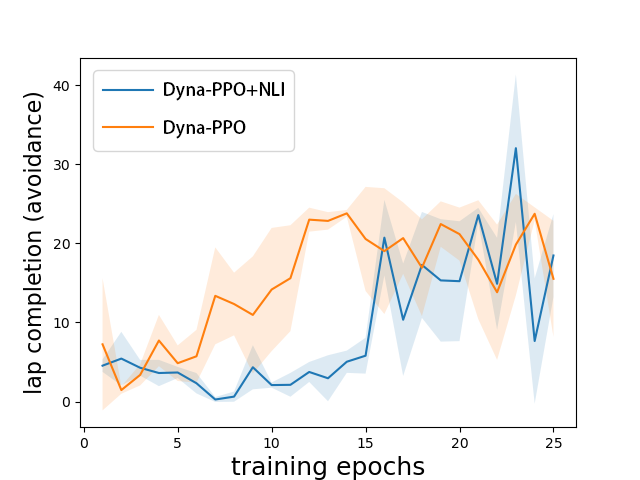}
        \label{fig:avoi_comp}
    \end{minipage}\hfill
    \begin{minipage}[t]{0.33\textwidth}
        \centering
        \includegraphics[width=0.98\linewidth]{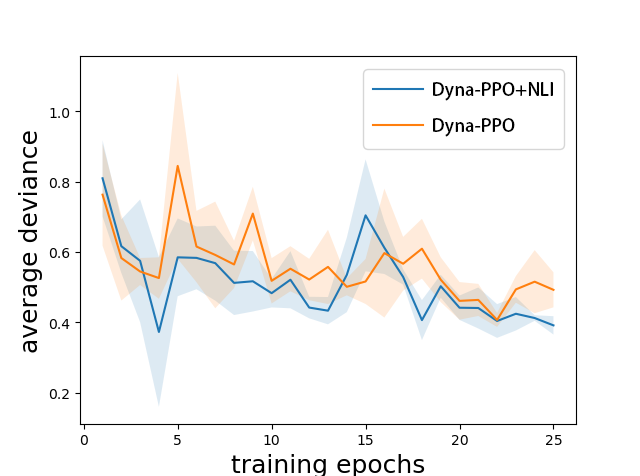}
        \label{fig:lane_dev}
    \end{minipage}\hfill
    \begin{minipage}[t]{0.33\textwidth}
        \centering
        \includegraphics[width=0.98\linewidth]{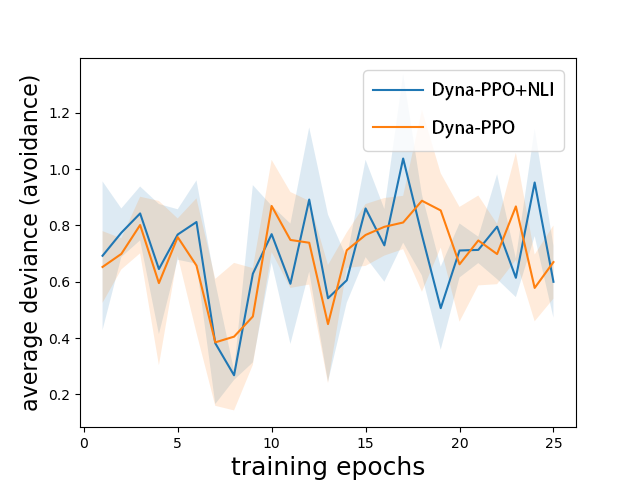}
        \label{fig:avoi_dev}
    \end{minipage}\hfill
    \caption{Learning curves comparison in different metrics}\label{learning_curve} 
\end{figure}

\subsection{Effects of using NL Voice Instructions}

To show the effectiveness of our methods with DRL, we test it on the full lap task. We compare the performance of the two models. One model is trained with NLI, and the other is not. The two models have the same hyper-parameter set and training/test environments. The Figure~\ref{learning_curve} shows the learning curves. In each training epoch, both models sample 3 trajectories for a round of gradient updates. The total training time is less than 1 hour for both models. We start coaching the model at 15 epochs, after which we can see Dyna-PPO+NLI's performance boosts, showing the effectiveness of NLI. The testing videos of the AV agent show the self-driving car smoothly turns with a handful of training episodes using NLI.
In fact, the two models' behaviors at the turning make the performance difference. 
The lap completion percentage of the Dyna-PPO agent does not increase after epoch 15 because it never successfully turns at the turning. The whole turning process takes about 30 to 60 successive steps, making it difficult for the random exploration strategy to find the correct steps.
However, this is not the case for Dyna-PPO+NLI. People can teach the agent the correct steps first. The agent then reinforces this skill because it leads to higher expected cumulative future rewards. This shows that the RL agent's random exploration strategy sometimes fails to learn optimal policy, and utilizing human knowledge for searching optimal policy during the training phase may help.

\section{Conclusion}

This paper proposes to use natural language instructions to help model-based RL when training autonomous driving cars. First, we collect voice instructions offline and build a training dataset and a taxonomy of NLIs. The data analysis shows that most natural language instructions are indeed about actions and rewards for the agent. Second, we build a model-based DRL framework that can incorporate live human voice instructions. Our method can allow both machine learning engineers and everyday end-users to 1) provide feedback to shape the reward function for various driving behaviors and 2) directly instruct the car agent to act. Experiment results show the effectiveness of our method. In the future, we plan to further improve the processing speed of speech-to-text and text classification due to the live nature of our work. Moreover, for actual deployment of the system, we must have a careful plan to handle real-time speech data in a privacy-preserving manner. Currently, speech recognition is done ``in the cloud'' and relies on an IT company's service. This can be improved when we develop highly effective in-house speech recognition tools. Through a human-friendly communication channel, this work presents a more natural way for AI agents to be trained and contributes to the bigger picture of human-robot cooperation.

\bibliographystyle{plainnat}
\bibliography{main}

\end{document}